# ORIGIN OF FERROELECTRICITY AND MULTIFERROICITY IN BINARY OXIDE THIN FILMS

*Maya Glinchuk[1], Anna Morozovska[2], and Lesya Yurchenko[1]*

[1] *Institute for Problems of Materials Science, National Academy of Sciences of Ukraine, Krjijanovskogo 3, 03142 Kyiv, Ukraine*

[2] *Institute of Physics, National Academy of Sciences of Ukraine,*

*46, pr. Nauky, 03028 Kyiv, Ukraine*

**Abstract**

The observation of ferroelectric, ferromagnetic and ferroelastic phases in thin films of binary oxides attract the broad interest of scientists and engineers. However, the theoretical consideration of observed behaviour physical nature was performed mainly for $HfO_2$ thin films from the first principles, and in the framework of Landau-Ginzburg-Devonshire (LGD) phenomenological approach with a special attention to the role of oxygen vacancies in both cases. Allowing for generality of the LGD theory we applied it to the group of binary oxides in this work. The calculations have been performed in the assumption that oxygen vacancies, as elastic dipoles, can be partially transformed into electric dipoles due to the defect site-induced and/or surface-induced inversion symmetry breaking (via e.g. piezoelectric effect), and can "migrate" entire the depth of an ultra-thin film. We calculated the dependence of the film polarization on the applied voltage for room temperature and different film thickness. Since the many films of binary oxide are ferroelectric and ferromagnetic due to the same oxygen vacancies, they can be multiferroics. Performed calculations have shown that thin films of binary oxides can be considered as a new wide class of multiferroics with broad spectra of physical properties useful for application in nanoelectronics and nanotechnology. The properties can be controlled by the choice of oxygen vacancies concentration, film thickness and special technological treatment, such as annealing.

# 1. Introduction

## A. Ferromagnetism in thin film of binary oxides

Investigation of magnetization of thin films of nonmagnetic in bulk oxides started in 2004, when Venkatesan et al. [1] reported about unexpected room-temperature magnetism in HfO$_2$ thin films on sapphire or silicon substrates. Then the observation of room-temperature magnetism in oxides thin films, such as HfO$_2$, TiO$_2$, SnO$_2$, and In$_2$O$_3$, on various substrates have attracted much attention of the scientific community to the so-called $d^0$-magnetizm [2, 3, 4, 5, 6, 7, 8].

The magnetic moment was measured with the help of SQUID magnetometer for the samples oriented perpendicular or parallel to applied magnetic field; at that parallel magnetic moment appeared to be 19 % smaller. Extrapolated Curie temperature was much higher than 400 K, magnitude of magnetic moment and hysteresis loop characteristics were shown to be dependent essentially on the type of substrate. These investigations, performed in Refs.[3, 4], allowed to pour light on the nature of magnetic defects in the thin films of oxides nonmagnetic in bulk, i.e. on the mechanisms of $d^0$ magnetism. Allowing for that all the films of HfO$_2$, TiO$_2$ and In$_2$O$_3$ were colorless, shiny and highly transparent, so that concentration of magnetic impurities can be well below $10^{-2}$ wt.%, the large value of magnetization is hard to attribute to any kind of impurities. Because of this it was supposed, that oxygen vacancies at the interface between the film and substrate are the main source of the magnetization (see e.g. Ref.[4, 9], where it was shown that oxygen vacancies are magnetic defects in vicinity of surface). The disappearance of magnetization after annealing in oxygen atmosphere [7] confirmed this supposition.

Another point that is important to emphasize is the strong thickness dependence of magnetic moment. Really, 10 nm thick TiO$_2$ and HfO$_2$ films magnetization of saturation was respectively 20 and 15 times larger than that of 200 nm thick films (see [3] and Refs. therein), i.e. it is approximately inversely proportional to the film thickness. Therefore, the number of magnetic defects (oxygen vacancies) in the film is independent on the thickness. One can directly interpret this as the evidence that the observed magnetization originates from defects localized mostly near the interface between the film and substrate. Allowing for that the physical reasons of the majority of abovementioned features stayed unexplained up to now, the Glinchuk et al [10] have found out that the oxygen vacancies localization mostly at the interfaces are the reasons of binary oxide films magnetization.

## B. Ferroelectricity in thin film of binary oxides

Recent observations of ferroelectricity in thin films of TiO$_2$, HfO$_2$, and related solid solutions has riveted the attention of scientific community both from fundamental perspective and due to potential for applications [11, 12, 13, 14]. For the latter, ferroelectricity in HfO$_2$ thin films enables application in ferroelectric memories due to ease of synthesis and compatibility with Si processing [15, 16, 17, 18]. From the application perspective, ferroelectricity in binary oxide thin films has a crucial relevance for performance of ferroelectric memories [15, 19], highly scalable and manufacturable ferroelectric materials of choice for advanced nonvolatile and random access memories [20].

Hoffmann et al [21] revealed that doping with Gd, that creates oxygen vacancies because of charge compensation necessity, induces ferroelectric phase in thin films of $HfO_2$, at the same time essential role of TaN electrodes due to interface oxidation of the electrodes was demonstrated. Pešić et al. [22] have found out that during the wake-up technological process of the device no new defects are generated but the existing defects including oxygen vacancies redistribute within the device. Different phases were revealed in thin films of $HfO_2$ doped with different amount of Si, including the ferroelectric phase [23], but corresponding phonon modes in $HfO_2$ [24] appeared slightly dependent on temperature in wide temperature range 20 – 800 K.

Allowing for all the above facts, vacancy diffusion has been identified as the main cause for the phase transformation and consequent increase of the remanent polarization in the binary oxide films [25]. Notably, that results of Ref. [25] for polarization and dielectric susceptibility hysteresis loops shape, remanent polarization, coercive field, as well as the changes of these characteristic taking place with the film thickness increase are in agreement with experimental loops measured by Polakowski and Muller [20].

Since the binary oxide film is ferroelectric and ferromagnetic due to the same oxygen vacancies, it can be a multiferroic, and this work explore the issue.

## 2. Thermodynamics of ferroelectricity appearance

This permits us to use LGD type free energy functional for quantitative consideration of ferroelectricity induced by **oxygen vacancies** in binary oxides. Gibbs potential density of oxide film with a paraelectric nonlinearity [26] has the following form [27],

$$G = G_V + G_S, \tag{1a}$$

$$G_V = \int_0^h d^3r \left( \begin{array}{l} \left( \dfrac{a_{ij}}{2} - Q_{klij}\sigma_{kl} \right) P_i P_j + \dfrac{a_{ijkl}}{4} P_i P_j P_k P_l + \dfrac{g_{ijkl}}{2} \dfrac{\partial P_i}{\partial x_j} \dfrac{\partial P_k}{\partial x_l} - P_i E_i(\mathbf{r}) \\ + \dfrac{F_{ijkl}}{2} \left( \sigma_{kl} \dfrac{\partial P_i}{\partial x_j} - P_i \dfrac{\partial \sigma_{kl}}{\partial x_j} \right) - \dfrac{s_{ijkl}}{2} \sigma_{ij}\sigma_{kl} - u_{ij}^W [\delta N_d(\mathbf{r})] \sigma_{ij} \end{array} \right), \tag{1b}$$

In Equation (1b) $P_i$ are the components of polarization vector ($i = 1, 2, 3$) and $\sigma_{ij}$ is the elastic stress tensor. The summation is performed over all repeated indices. Equations of state $\partial G/\partial \sigma_{ij} = -u_{ij}$ determine the strains $u_{ij}$. Euler-Lagrange equations $\partial G/\partial P_i = 0$ determine the polarization components. The surface energy $G_S$ is defined by the oxide-substrate and oxide-electrodes interface chemistry.

The expansion coefficients $a_{ij}$ are positive and typically diagonal, $a_{ij} = \delta_{ij} a_{ii}$, because the ferroelectricity is absent in the absence of vacancies. The coefficient $a_{ii}$ is related with a bulk relative dielectric permittivity tensor $\varepsilon_{ij}$ as $a_{ii} = \dfrac{1}{\varepsilon_0 (\varepsilon_{ii} - 1)}$, where $i = 1, 2, 3$; $\varepsilon_0$ is a universal dielectric constant and $\varepsilon_{ii} \approx 25$ for $HfO_2$ or $\varepsilon_{ii} \approx 80$ for $TiO_2$ [28]. The nonlinearity $\beta = a_{33}$ can be estimated for $HfO_2$ as $5 \times 10^{10}$ J m$^5$/C$^4$.

The matrix of the gradient coefficients $g_{ijkl}$ is positively defined. $Q_{ijkl}$ is the electrostriction tensor, $s_{ijkl}$ is the elastic compliances tensor, $F_{ijkl}$ is the forth-rank tensor of flexoelectric coupling ($F_{ijkl} \sim 10^{-11}$ C$^{-1}$m$^3$). $E_i(\mathbf{r})$ denotes the electric field.

The last term in Eq.(1) includes a Vegard-type concentration-deformation energy, $u_{ij}^{W}[\delta N_d(\mathbf{r})]\sigma_{ij}$, determined by the random mobile vacancies (charged or electroneutral) with concentration $\delta N_d(\mathbf{r}) \sim \left\langle \sum_k \delta(\mathbf{r} - \mathbf{r}_k) \right\rangle - \overline{N}_{3D}$. The vacancies concentration in the bulk is small enough in thermodynamic equilibrium, i.e. $\overline{N}_{3D} \ll 10^{28}$ m$^{-3}$. The vacancies, or their complexes, tend to accumulate in the vicinity of any inhomogeneities, surfaces and interfaces [see **Fig.1(a)**], since the energy of their formation in such places, as a rule, is much smaller than in the homogeneous bulk [29, 30, 31, 32]. The vacancies can create sufficiently strong fields in the places of their accumulation, which in turn can lead to new phases appearance in oxides, for example, polar (ferroelectric) ones. On contrary, the non-polar state remains in the places where there are few vacancies. So, the polar ferroelectric and nonpolar states coexistence can be realized in the films.

The average distance between defects centres $2R$ should be associated with average size per inclusion. The defect size $r_0$ is much smaller than $R$, e.g. $r_0 \ll R$, where $r_0$ is ionic radius $\sim (0.1 - 1)$Å$^3$ [see **Fig.1(b)**].

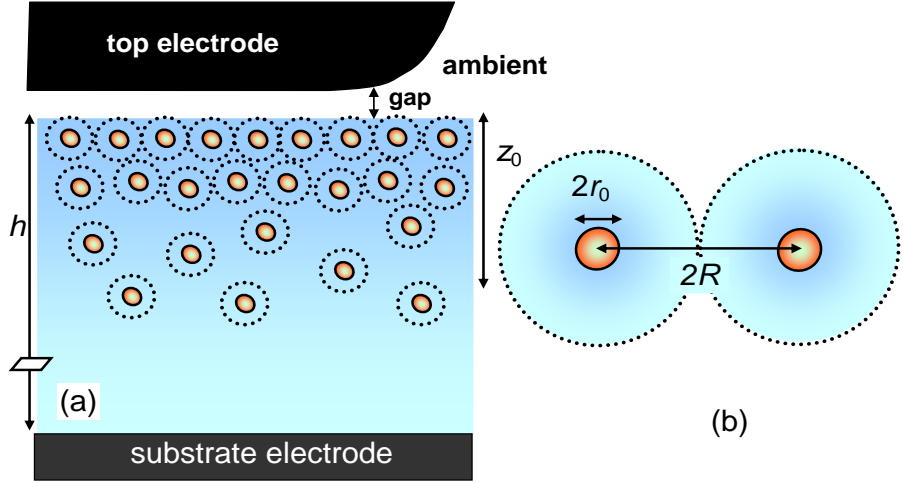

**FIGURE 1. (a)** Defect accumulation near the film surface. **(b)** Schematics of the spherical defects with radius $r_0$ embedded in the film of thickness $h$. The average distance between two defects is at the surface is $2R$, $r_0 < R$. For a defect at the surface $\bar{N}_{2D} = \dfrac{1}{4\pi R^2}$.

A dilatation center with equal distortion can be considered as a simple elastic model of impurity atom or vacancy, whose own distortion (Vegard strain) is [33, 34]

$$w_{xx}^k(\mathbf{r}) = w_{yy}^k(\mathbf{r}) = w_{zz}^k(\mathbf{r}) = W\delta(\mathbf{r} - \mathbf{r}_k) \tag{2}$$

Equations (2) are derived under the assumption of isotropic and diagonal Vegard expansion tensor $W_{ij}$, $W_{ij} = W\delta_{ij}$, where $W = \pm(5 - 20)$ Å$^3$ [35] is the volume change in the point of defect localization $\mathbf{r} = \mathbf{r}_k$. In general case the structure of Vegard expansion tensor $W_{ij}$ (elastic dipole) is controlled by the symmetry (crystalline or Curie group symmetry) of the material [36, 37, 38].

Nonzero components of the elastic displacement, strain and stresses induced by a spherically-symmetric elastic point defect (e.g. dilatation centre) located in the coordinate origin, $\mathbf{r} = 0$, have the form [33-34]:

$$\mathbf{u}^W(\mathbf{r}) = -\frac{(1+\nu)W}{4\pi(1-\nu)}\frac{\mathbf{r}}{r^3} \equiv \frac{(1+\nu)W}{4\pi(1-\nu)}\nabla\left(\frac{1}{r}\right), \tag{3a}$$

$$u_{ii}^W(\mathbf{r}) = -\frac{(1+\nu)W}{4\pi(1-\nu)}\frac{\partial}{\partial x_i}\left(\frac{x_i}{r^3}\right) = -\frac{(1+\nu)W}{4\pi(1-\nu)}\frac{r^2 - 3x_i^2}{r^5} \equiv \frac{(1+\nu)W}{4\pi(1-\nu)}\frac{\partial^2}{\partial x_i^2}\left(\frac{1}{r}\right), \tag{3b}$$

$$u_{ij}^W(\mathbf{r}) = -\frac{(1+\nu)W}{4\pi(1-\nu)}\frac{\partial}{\partial x_i}\left(\frac{x_j}{r^3}\right) = \frac{(1+\nu)W}{4\pi(1-\nu)}\frac{3x_i x_j}{r^5} \quad (i \neq j) \tag{3c}$$

$$\sigma_{ii}^W(\mathbf{r}) = -\frac{G(1+\nu)W}{2\pi(1-\nu)}\frac{r^2 - 3x_i^2}{r^5} \equiv \frac{G(1+\nu)W}{4\pi(1-\nu)}\frac{\partial^2}{\partial x_i^2}\left(\frac{1}{r}\right), \tag{3d}$$

$$\sigma_{ij}^W(\mathbf{r}) = -\frac{G(1+\nu)W}{2\pi(1-\nu)}\frac{3x_i x_j}{r^5} \equiv -\frac{G(1+\nu)W}{4\pi(1-\nu)}\frac{\partial}{\partial x_i}\left(\frac{x_j}{r^3}\right) \quad (i \neq j). \tag{3e}$$

The radius $r = \sqrt{x_1^2 + x_2^2 + x_3^2}$ is introduced in Eqs.(3). The Poisson ratio is $v = -s_{12}/s_{11}$ for cubic m3m symmetry. $G$ is the shear modulus and $s_{ij}$ are elastic compliances.

Substitution of elastic fields (3) into the potential Eq.(1b) leads to the renormalization of the coefficient $a_{ij} \to \alpha_{ij}^R$ by the electrostriction coupling with Vegard expansion. For diagonal component this gives:

$$\alpha_{kk}^R(\mathbf{r}) \cong \frac{1}{\varepsilon_0(\varepsilon_{kk}-1)} - 2Q_{ijkk} \sum_l \sigma_{ij}^W [\delta N_d(\mathbf{r}-\mathbf{r}_l)] \quad (4)$$

The summation is performed over all vacancies sites. One can see from Eq.(4) that the local polar state, occurring under the condition $\alpha_{kk}^R < 0$, is not excluded in the spatial regions, where the defects and carriers concentration is high enough.

## A. Soft mode origin of ferroelectricity

Using ergodic hypothesis the averaging and summation in Eq.(4) reduces to the averaging over the film volume. The averaging of the function $2Q_{ij33}\sigma_{ij}^W[\delta N_d(\mathbf{r})]$ in Eq.(4) reduces to the averaging over the cylindrical volume, $4\pi R^2 h$, and for m3m symmetry it gives

$$\langle \alpha_{ii}^R(\mathbf{r}) \rangle = \frac{1}{\varepsilon_0(\varepsilon_{ii}-1)} + 2(Q_{ii} - Q_{ij}) \int_0^h dz \int_0^R \rho d\rho \frac{G(1+v)W}{2\pi(1-v)} \frac{r^2 - 3x_i^2}{r^5} f(\mathbf{r}) \quad (i=1, 2, 3) \quad (5)$$

Here $\rho = \sqrt{x_1^2 + x_2^2}$ is the polar radius, $z = x_3$, and we used that $Q_{11} = Q_{22} = Q_{33}$ and $Q_{12} = Q_{13} = Q_{23}$ for m3m symmetry; and the identity $2r^2 - 3(x_1^2 + x_2^2) \equiv 3x_3^2 - r^2$. The film thickness is $h$. The average 2D-concentration of defects is $\overline{N}_{2D} = \frac{1}{4\pi R^2}$. The distribution of vacancies depends on the distance from the film surface, and reveals exponential or linear decay, namely [39]:

$$f(\mathbf{r}) = \frac{1}{\pi R^2 h} \exp\left(-\frac{z}{z_0}\right) \approx \begin{cases} \frac{1}{4\pi R^2 h}\left(1 - \frac{z}{z_0}\right), & 0 < z < z_0, \\ 0, & z > z_0. \end{cases} \quad (6)$$

Here $z_0$ is the decay factor of defect concentration under the surface. Typically $z_0 \ll h$. Substitution of Eq.(6) in Eq.(5) yields

$$\langle \alpha_{33}^R(\mathbf{r}) \rangle = \frac{1}{\varepsilon_0(\varepsilon_{33}-1)} - 2(Q_{33} - Q_{13})\frac{1+v}{1-v} GW \int_0^h \frac{dz f(z) R^2}{\sqrt{(R^2+z^2)^3}}$$

$$\approx \frac{1}{\varepsilon_0(\varepsilon_{33}-1)} - (Q_{33} - Q_{13})\frac{1+v}{1-v}\frac{2GW}{\pi R^2 h}\left(\frac{z_0}{\sqrt{R^2+z_0^2}+R} - 1\right) \quad (7a)$$

$$\langle\alpha_{11}^R(\mathbf{r})\rangle = \langle\alpha_{22}^R(\mathbf{r})\rangle = \frac{1}{\varepsilon_0(\varepsilon_{11}-1)} - 2(Q_{11}-Q_{12})\frac{1+v}{1-v}GW\int_0^h \frac{dz f(z)z^2}{3\sqrt{(R^2+z^2)^3}}$$

$$\approx \frac{1}{\varepsilon_0(\varepsilon_{11}-1)} - (Q_{11}-Q_{12})\frac{1+v}{1-v}\frac{2GW}{3\pi R^2 h}\left(\frac{2z_0}{\sqrt{R^2+z_0^2}+R} - \text{Arcsinh}\left(\frac{z_0}{R}\right)\right), \quad (7b)$$

The approximate equality in Eq.(6) is valid under the validity of the inequality $z_0 \ll h$.

Using HfO$_2$ parameters $\beta = 5\times10^{10}$ J m$^5$/C$^4$, $G = 109$ GPa, $v = 0.3$ [40], $\varepsilon_{ii} \approx 25$ and electrostriction coefficients $Q_{33} = 0.7$ m$^4$/C$^2$, $Q_{12} = -0.5$ m$^4$/C$^2$, $R = 2$ nm, $z_0 = 5$ nm, $h = 10$ nm and Vegard coefficient $W = -20$ Å$^3$ we obtained that the first term in Eq.(6), $\frac{1}{\varepsilon_0(\varepsilon_{33}-1)} \approx 4\times10^9$ m/F, is at least two orders of magnitude higher than the second one induced by the electro-chemical coupling. So that $\langle\alpha_{33}^R(\mathbf{r})\rangle \approx \frac{1}{\varepsilon_0(\varepsilon_{33}-1)} > 0$ is always positive and corresponding spontaneous polarization $P_S = \sqrt{-\langle\alpha_{33}^R(\mathbf{r})\rangle/\beta}$ does not exist in the sense of mean field theory. A qualitatively similar estimate is valid for TiO$_2$, but here the difference is much smaller. In what follows we will show that the contribution of flexoeffect and Vegard stress will lead to the ferroelectricity appearance.

**B. Flexo-chemical origin of ferroelectricity**

Next let us estimate the polarization and electric fields variations induced by the joint action of Vegard stresses and flexoelectric coupling. Equations of state $\partial G/\partial\sigma_{ij} = -u_{ij}$ give the strains $u_{ij}$ as:

$$u_{ij} = s_{ijkl}\sigma_{kl} + \sum_l u_{ij}^W[\delta N_d(\mathbf{r}-\mathbf{r}_l)] - F_{ijkl}\frac{\partial P_l}{\partial x_k} + Q_{ijkl}P_k P_l. \quad (7)$$

In linear approximation polarization variation $\delta P_i$ is induced by defects (2) located at $\mathbf{r} = \mathbf{r}_m$ with elastic fields (3) due to the joint action of Vegard stresses and flexoelectric coupling. In particular the variation $\delta P_i$ can be estimated as $\delta P_i(\mathbf{r}) = \tilde{f}_{ijkl}\frac{\partial u_{jk}^W[\delta N_d(\mathbf{r})]}{\partial x_l}$. For z-component this gives nonzero average:

$$\delta P_3(\mathbf{r}) \approx \frac{(1+v)W}{4\pi(1-v)}\sum_m\left(\tilde{f}_{33}\frac{\partial^3}{\partial x_3^3} + \tilde{f}_{31}\left(\frac{\partial^3}{\partial x_3\partial x_1^2} + \frac{\partial^3}{\partial x_3\partial x_2^2}\right)\right)\frac{1}{|\mathbf{r}-\mathbf{r}_m|}, \quad (8a)$$

$$\langle\delta P_3(\mathbf{r})\rangle = \frac{(1+v)W}{2(1-v)}(\tilde{f}_{33}-\tilde{f}_{31})\int_0^h dz f(z)\frac{\partial^3}{\partial z^3}\sqrt{R^2+z^2} \approx \frac{1+v}{1-v}W\frac{(\tilde{f}_{33}-\tilde{f}_{31})}{2\pi R^2 h}\left(\frac{1}{\sqrt{R^2+z_0^2}}-\frac{1}{R}\right), \quad (8b)$$

Where $\tilde{f}_{ijkl}$ is the flexoelectric effect stress tensor related with the tensor $f_{ijkl}$ ~ (5–10) V by expression $\tilde{f}_{ijkl} = \langle \alpha_{33}^R(\mathbf{r}) \rangle f_{ijkl} \cong \varepsilon_0 (\varepsilon_{33} - 1) f_{ijkl}$. The value $\tilde{f}$ ~ $10^{-8}$ C/m can be estimated from Kogan model [41]. Estimates based on Eq.(8) gives relatively high polarization ~ (2–5) µC/cm² typical for improper ferroelectricity, and the hysteresis loops measured by Nishimura et al in thin undoped 20 nm $HfO_2$ films correspond to the same order of remanent polarization (~ 2.5 µC/cm²).

Polarization $P_S(T,h,N_S)$ dependences on the applied voltage calculated for room temperature, fixed $N_S$ and different film thickness $h$ = (5–25) nm are shown in **Fig. 2**. The loop width strongly decreases in with $h$ increase from 5 to 25 nm and the remanent polarization changes from 7 µC/cm² for $h$ = 5 nm to 1 µC/cm² for $h$ = 25 nm, and the loop becomes rather slim and tilted with $h$ increase from 5 nm to 25 nm [compare loops 1 – 5 in **Fig. 2**]. Further increase of the film thickness $h$ does not lead to any significant changes in the loop shape and sizes, because for $h$ higher than 25 nm the loop itself is the dynamic effect.

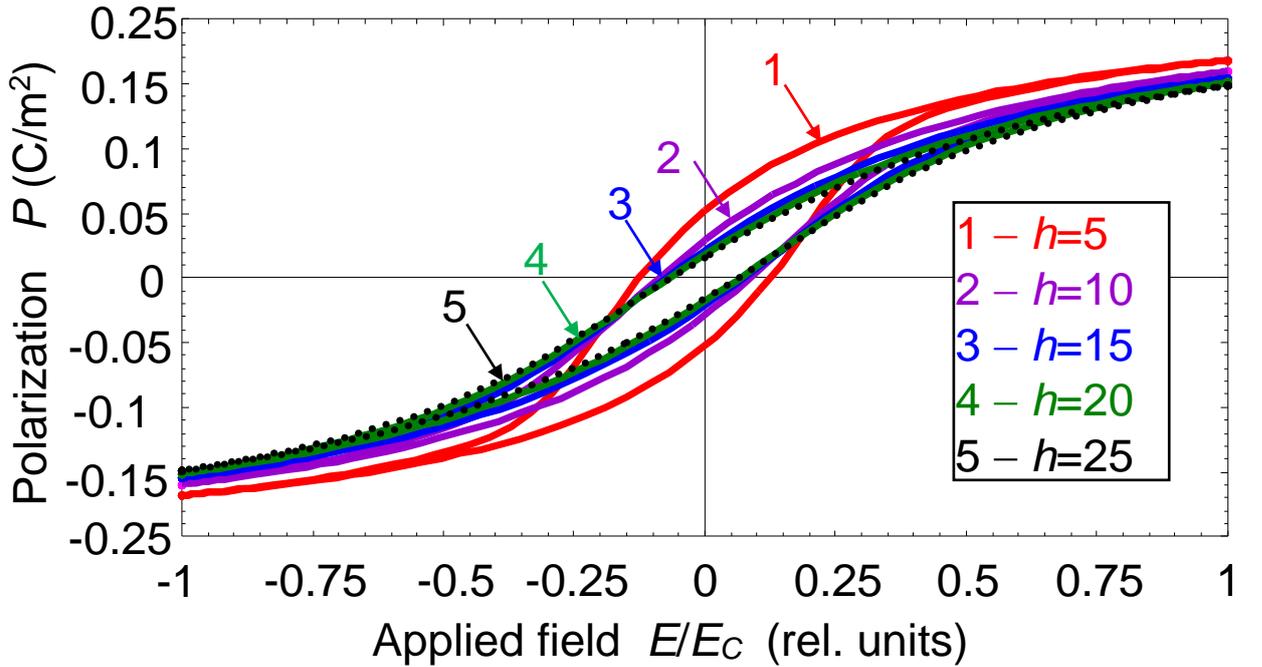

**FIGURE 2.** Polarization $P_S(T,h,N_S)$ dependences on applied voltage calculated for room temperature $T$ = 298 K, $N_S$ = $10^{25}$ m⁻³ and different $h$ = (5–25) nm (curves 1, 2, 3, 4, 5). Other parameters are the same as in Ref. [25].

## C. The origin of multiferroicity

Since the binary oxide films are ferroelectric and ferromagnetic due to the same oxygen vacancies, they can be considered as a new group of multiferroics. Moreover, first principle calculations and their comparison with experiment had shown [21] that oxygen vacancies as elastic dipoles can lead to appearance of structural ferroelastic phases also.

Generally speaking, oxygen vacancies in binary oxides thin films can be considered as the materials with a mixture of electric and elastic dipoles. By definition oxygen vacancies are known to be elastic dipoles, which transform into electric dipoles in vicinity of surfaces due to piezoelectric effect. Indeed, the contribution of elastic dipoles into total energy $\mathcal{E}_{tot}$ can be written as $\mathcal{E}_{tot} = w_{ij}u_{ij}$, where $w_{ij}$ and $u_{ij}$ are respectively elastic dipole and mechanical tension. This tension due to piezoelectric effect leads to appearance of electric field $E_k = u_{ij}\eta_{ijk}$, where $\eta_{ijk}$ is piezoelectric coefficient tensor. The contribution of electric field into energy can be written as its product with electric dipole $d_k$. Substituting of $u_{ij} = E_k/\eta_{ijk}$ into above written expression for energy one obtains $\mathcal{E}_{tot} = w_{ij}E_k/\eta_{ijk} = d_k E_k$, so that electric dipole moment of oxygen vacancy $d_k = w_{ij}/\eta_{ijk}$ and its elastic dipole moment $w_{ij} = d_k\eta_{ijk}$. It is obvious that phase diagram of the oxides thin films has to depend on general concentration of oxygen vacancies to overcome the percolation threshold for appearance of long-range order. The first principle calculations of HfO$_2$ thin film phase diagram as function of oxygen vacancies concentrations had shown the appearance of orthorhombic FE phase, cubic, tetragonal, monoclinic paraelectric (ferroelastic) phases at room temperature.

Therefore, binary oxides thin films indeed can be considered as new wide class of multiferroics with coexistence of ferroelectric, ferroelastic and magnetic phases, and thus with broad spectra of physical properties useful for applications in modern nanoelectronics and nanotechnology. The choice of necessary properties can be controlled by the concentration of oxygen vacancies, films thickness and special technological procedure.

## IV. DISCUSSION AND CONCLUSION

The calculations have been performed in the assumption that oxygen vacancies, as elastic dipoles, can be partially transformed into electric dipoles due to the defect site-induced and/or surface-induced inversion symmetry breaking (via e.g. piezoelectric effect), and can "migrate" entire the depth of an ultra-thin film.

The proposed model opens the way for calculation of binary oxide films multiferroic properties and complex phase diagrams. To perform the calculations of the phase diagram one has to consider polarization and mechanical tension induced by oxygen vacancies, electric and elastic dipoles, respectively, as long-range order parameters of ferroelectric and ferroelastic phase transitions. Keeping in mind the existence of magnetization in many binary oxide thin films it can be considered as the third long-range order parameter. Therefore magnetoelectric and magnetoelastic interaction will represent the coupling between magnetization and two other order parameters, and the coupling between them being represented by Vegard-type interaction.

The developed theory opens the way for the selection of binary oxide films optimal properties (namely thickness, vacancies concentration and annealing time) to obtain high performance electronic devices based on highly scalable, Si-compatible and manufacturable ferroelectric thin dioxide films for ferroelectric memories and capacitor industry.

Allowing for a generality of the proposed approach, it can be applied to binary oxides, both undoped and doped ones. Furthermore, it is applicable to other cases of dipolar impurities, originated from the trapped charges at the interfaces, etc.

**Acknowledgments.** A.N.M. work was partially supported by the European Union's Horizon 2020 research and innovation program under the Marie Skłodowska-Curie (grant agreement No 778070).